\documentstyle[12pt]{article}
\begin{document}
\title{A NOTE ON DEGENERATE AND ANAMOLOUS BOSONS}
\author{B.G. Sidharth\\
Centre for Applicable Mathematics \& Computer Sciences\\
B.M. Birla Science Centre, Adarsh Nagar, Hyderabad - 500 463\\
India}
\date{}
\maketitle
\begin{abstract}
In this note it is shown that for a mono-energetic collection of
Bosons, at a certain (non-zero) momentum or temperature, there is
condensation while there is another momentum or temperature at
which there is infinite dilution and below which the gas exhibits
anamolous Fermionic behaviour.
\end{abstract}

We start with the well known formula for the occupation number in a
Bose gas\cite{Huang}:
\begin{equation}
\langle n_p\rangle = \frac{1}{z^{-1}e^{\beta \epsilon_p} -1}\label{e1}
\end{equation}
where
$$\beta \equiv  1/kT.$$
We also define,
$$z' \equiv \lambda^3/v \equiv zb, \lambda =
\left({\frac{2\pi \hbar^2}{mkT}}\right)^{1/2}$$

When $\epsilon_p = 0, z=1$, we have from (\ref{e1}), $\langle n_0\rangle =
\infty$.
This is the Bose-Einstein condensation.

Let us now consider a mono-energetic collection of Bosons:
\begin{equation}
\langle{n_p}' \rangle = \langle n_p\rangle \delta(p-p_0)\label{e2},
\end{equation}
so that
\begin{equation}
kT \approx \epsilon_{p_0} = \frac{p_0^2}{2m}, \quad
\lambda = (4\pi)^{1/2} \frac{\hbar}{p_0},\quad \beta \epsilon_{p_0} \approx
1\label{e3}
\end{equation}
{}From (\ref{e2}) we get,
\begin{equation}
N = \frac{V}{\hbar^3} \int_0^\infty 4\pi p^2dp\delta
(p-p_0)\langle n_p\rangle,\label{e4}
\end{equation}
where $N$ is the total number of particles and $V$ is the total volume. Using
(\ref{e1}) and (\ref{e3}) in (\ref{e4}) we get,
$$1 = \frac{4\pi v p_0^2}{\hbar^3} \langle n_{p_0}\rangle =
\left[{\frac{(4\pi )^{5/2}}{p_0}}\right]
\frac{v}{\lambda^3} \left[{\frac{1}{z^{-1} e-1}}\right]$$
so that we have
\begin{equation}
z'^{-1} \left[{\frac{(4\pi )^{5/2}}{p_0}- eb}\right] = -1\label{e5}
\end{equation}
We can see from (\ref{e5}) that if
\begin{equation}
p_0 \approx \frac{(4\pi )^{5/2}}{eb-1}\label{e6}
\end{equation}
(or at the corresponding temperature) then $z'\approx 1$. But this is
the condition for condensation: Remembering that $\lambda$ is of the
order of the particles' deBroglie wave length and $v$ is the average
volume per particle, this means that the gas gets very densely packed.
On the other hand, if
\begin{equation}
p_0 \approx \frac{(4\pi )^{5/2}}{eb}\label{e7}
\end{equation}
then as can be seen from (\ref{e5}) $z' \approx 0$. In this case we have
the opposite effect: The gas becomes very dilute.
Finally, if
\begin{equation}
p_0 <\frac{(4\pi )^{5/2}}{eb},\label{e8}
\end{equation}
then equation (\ref{e5}) leads to a contradiction: We require that
$z' <0$, which is not possible.

The contradiction disappears if we realize that for momenta given by
(\ref{e8}) or for the corresponding temperatures the Bosons effectively
behave like Fermions. In this case, the average occupation number is
given, instead of (\ref{e1}), by,
$$\langle n_p\rangle = \frac{1}{z^{-1}e^{\beta \epsilon_p}+1}$$
and instead of (\ref{e5}) we have,
$$z'^{-1} \left[{\frac{(4\pi)^{5/2}}{p_0}- eb}\right] = +1$$
and condition(\ref{e8}) poses no problem.
We now characterize $b$, which we have defined above in $z'=zb$. It
is known that (cf.ref.\cite{Huang}),
$$z'\left[{1 - \frac{\langle n_0\rangle}{N}}\right] = g_{3/2}(z)$$,
where $\langle n_0\rangle$ is the occupation number for energy or
momentum $=0$. As we are dealing with a mono-energetic Bose gas with
non-zero energy (i.e. at non-zero temperature), we have,
$$\frac{\langle n_0\rangle}{N} \approx 0,$$
so that,
\begin{equation}
z' \equiv zb \approx g_{3/2}(z) = z\left[{1+\frac{z}{2^{3/2}}+
\frac{z^2}{3^{3/2}}+\ldots}\right] \quad (0 \le z \le 1)\label{e9}
\end{equation}
It easily follows that (cf.ref\cite{Huang}),
$$1 < b < 2.6$$.
Infact if $z<<1$, we have from (\ref{e9}),
\begin{equation}
b \approx 1\label{e10}
\end{equation}
This is the case (\ref{e7}) where $z'$ and consequently $z$ are $\approx 0$.
So (\ref{e7}) now becomes,
\begin{equation}
p_0 \approx \frac{(4\pi)^{5/2}}{e}\label{e11}
\end{equation}
Coming now to the inequality (\ref{e8}), we observe that for Fermions,
we have instead of (\ref{e9}), (cf.ref.\cite{Huang}),
\begin{equation}
z'\equiv zb = f_{3/2}(z) = z - \frac{z^2}{2^{3/2}}+
\frac{z^3}{3^{3/2}}\label{e12}
\end{equation}
If now the inequality (\ref{e8}) is nearly an equality, as we can choose, then,
as
in the case of (\ref{e7}), we will have, $z'\approx 0$ so that
$z \approx 0$ also and from (\ref{e12}) we can see that $b$ satisfies
(\ref{e10}). So the inequality (\ref{e8}) becomes,
\begin{equation}
p_0 < \frac{(4\pi)^{5/2}}{e}\label{e13}
\end{equation}
We finally come to the case of condensation given by (\ref{e6}). Here,
from (\ref{e9}) it follows that
$$g_{3/2}(z) \approx 1,$$
which gives,
$$z \approx 0.7, b\approx 1.4.$$
Inserting this value of $b$ in (\ref{e6}), we get,
\begin{equation}
p_0 \approx \frac{(4\pi)^{5/2}}{1.4e{-1}}\label{e14}
\end{equation}
To sum up, for a collection of mono-energetic Bosons, ({\it i}) at the momentum
given by (\ref{e14}) we have condensation; ({\it ii}) at the momentum given
by (\ref{e11}) we have infinite dilution; ({\it iii}) for momenta given by
(\ref{e13}) we have anamolous Fermionic behaviour.

\end{document}